\documentstyle[a4,12pt]{article}

\begin{document}

\author{Alexander M. Lazaruk, Nikolay V. Karelin \\ \\
Institute of Physics, Belarus Academy of Sciences,\\
Minsk 220072, Belarus
\footnotemark[1]}
\title{\bf Average Number of Coherent Modes for Pulse Random Fields}
\date{}
\maketitle
\footnotetext[1]{Further author information --\\
A. M. L.: E-mail: lazaruk@bas33.basnet.minsk.by; 
Telephone: (375) 017-268-4419; Fax: (375) 017-239-3131\\
N. V. K.: E-mail: karelin@bas33.basnet.minsk.by}

\begin{abstract}
Some consequences of spatio-temporal symmetry for the deterministic
decomposition of complex light fields into factorized components are
considered. This enables to reveal interrelations between spatial and
temporal coherence properties of wave. An estimation of average number of 
the decomposition terms is obtained in the case of statistical ensemble
of light pulses.
\end{abstract}


\section*{1. INTRODUCTION}

Modal description of light coherence, being a multidimensional
generalisation of the well-known Karhunen-Lo\'eve expansion, was first
introduced in optics by Gamo\cite{Gamo}. In short, the essence of the
approach lies in the fact that any correlation function of a field $E({\bf r},t)$
 --- in particular the transverse beam coherence 
$\Gamma _S({\bf r},{\bf r'})$ --- can be expressed as a superposition of 
factorized components 
$$
\Gamma _S({\bf r},{\bf r'})=\int dtE({\bf r},t)E^{*}({\bf r'},t)
=\sum_nu_n{\cal E}_n({\bf r})\,{\cal E}_n^{*}({\bf r'}),
\eqno(1)
$$
where each term of the sum represents a completely coherent partial wave 
${\cal E}_n({\bf r})$. The decomposition basic functions and modal energies 
$u_n$ are eigenvectors and eigenvalues of Fredholm's integral equation 
$$
u_n\,{\cal E}_n({\bf r})=\int d^2r'\,\Gamma _S({\bf r},{\bf r}%
')\,{\cal E}_n({\bf r}')\eqno(2)
$$
and, since the kernel of (2) is Hermitian, the set of functions ${\cal E}_n(%
{\bf r})$ is orthonormal 
$$
\int d^2r{\cal E}_n({\bf r}){\cal E}_m^{*}({\bf r})=\delta _{n,m}.
\eqno(3)
$$

In (1) the transverse correlation function $\Gamma _S({\bf r},{\bf r'})$ 
is determined as a time average (over the pulse duration or
time of registration) and, in this sense, it is a deterministic
characteristic of any particular wave. In the case of statistical ensemble
of similar pulses the time averaging can be replaced by statistical one or
both types of averaging may be combined. Formally it has no effect on
relations (1) -- (3), but, as it will be seen later, changes their physical
meaning.

The modal expansion (1) -- (3), as well as its modification for
space-frequency domain\cite{Wolf1,Wolf2}, is broadly used in 
coherence theory as a convenient tool for estimation of entropy and 
informational capacity of light beams\cite{Gamo}, as the best method for 
modelling of coherence properties of a complex light wave with a finite set of 
simple mutually incoherent waves\cite{DeSantis}, and so on. These relations 
constitute the mathematical basis for proof of various types of uncertainty 
inequalities\cite{Bastiaans}.

The next principal step in the development of the modal formalism was made
by Pasmanik and Sidorovich\cite{Pasm}. They demonstrated the 
spatio-tempo\-ral
symmetry for decomposition (1) -- (3), that leads to some important
relations between spatial and temporal coherence characteristics of light
waves. The discussion of them both in deterministic form and under application
of the ensemble averaging is a main goal of the present paper.

\section*{2. DUAL APPROACH OVERVIEW}

Let us suppose that for some wavefield $E({\bf r}, t)$, where ${\bf r}$ is a 
two-dimensional radius-vector at a plane $z=const$, the modal expansion 
(1)~--~(3) is known. So far the solutions of the integral equation (2) form 
a complete functional basis (when including the functions, corresponding to 
zero eigenvalues\cite{Lumley}), one can define a set of projections of the 
initial field onto this basis 
$$
\sqrt{u_n}\,e_n(t)=\int d^2r\,E({\bf r},t)\,{\cal E}_n^{*}({\bf r}),
\eqno(4) 
$$
that, in turn, allows to build up a modal representation of the field
itself 
$$
E({\bf r},t)=\sum_n\sqrt{u_n}{\cal E}_n({\bf r})\,e_n(t). 
\eqno(5) 
$$

The most important point here is that for pulses of finite total energy 
$$
U=\int d^2r\int dt\left| E({\bf r},t)\right| ^2<\infty 
$$
the projections $e_n(t)$ also constitute the complete orthonormal set of
basic functions 
$$
\int dt\, e_n(t) e_m^{*}(t)=\delta _{n,m}. 
\eqno(6)
$$
Last relation can be proved by direct substitution of definition (4) into
(6) and accounting (3). It is just this mutual orthogonality of temporal
functions $e_n(t)$, that leads to absence of any interference between
different terms of spatial basis.

Another approach to evaluation of temporal basis (4) lies in use of dual
integral equation 
$$
u_n\,e_n(t)=\int dt'\,\Gamma _T(t,t')\,e_n(t'),
\eqno(7)
$$
where $\Gamma _T(t,t')$ is a global temporal correlation function of the
field $E({\bf r},t)$ 
$$
\Gamma_T (t,t')=\int d^2r\,E({\bf r},t)\,E^{*}({\bf r},t')=
\sum_nu_n\,e_n(t)\,e_n^{*}(t').
\eqno(8)
$$
In contrast to the standard definition\cite{Wolf1} (with averaging over the 
time or ensemble of pulses) the averaging procedure in (8) is carried out 
over the beam cross-section. Hence, the function $\Gamma_T(t,t')$ expresses 
the overall correlation the wavefront patterns and is closely related to the 
degree of similarity\cite{Lazaruk,ZeldShkun} $H(t,t')$ of the wavefield for 
consequent time moments 
$$
H(t,t')=\left| \int d^2r\,E({\bf r},t)\,E^{*}({\bf r},t')\right| ^2\,
\bigg/\left( \int d^2r\,\left| E({\bf r},t)\right| ^2\right)
\left( \int d^2r\,\left| E({\bf r},t')\right| ^2\right) .
$$

One can easily see that equations (2) and (7) make up two {\sl equivalent dual} 
variants for evaluation the decomposition (5). Both equations have identical 
spectra of eigenvalues and for complete description of modal structure of 
field one needs to know only one set of basic functions ${\cal E}_n({\bf r})$
or $e_n(t)$. The second can be immediately determined through projection (4)
or via its dual equivalent 
$$
\sqrt{u_n}\,{\cal E}_n({\bf r})=\int dt\,E({\bf r},t)\,e_n^{*}(t).
\eqno(9)
$$
The last variant (6) -- (9) has an advantage of dealing with 1-D task.
One more exact consequence of the dual formalism is that two 
functions $\Gamma _S({\bf r},{\bf r'})$ and 
$\Gamma _T(t,t')$ in the case of no degeneracy (all $u_n$ are different) 
allow one to completely reconstruct the form of field.
Under degeneration (e.~g. when $u_i = u_j$) 
the ambiguity arises from the fact that two different wave structures 
$e_i(t){\cal E}_i({\bf r}) + e_j(t){\cal E}_j({\bf r})$ and 
$e_i(t){\cal E}_j({\bf r}) + e_j(t){\cal E}_i({\bf r})$ produce the same 
correlation functions (1), (8).

Strictly speaking, the modal structure of the field does not remain
constant under the wave propagation, but mode mixing is comparatively low
for quasimonochromatic beams with small divergence\cite{Pasm}. That is why
the deterministic dual decomposition (5) is inherently aimed to description
of laser pulses and has been first applied in nonlinear 
optics\cite{ZeldShkun}, where the partial coherence just means a high 
complexity of interacting waves.

In practice the complete modal description can be fulfilled only 
for very few classes of models\cite{StWol,Gori}, what is, first of all, 
related with intricacy of integral equations (2), (7) solving. Therefore those 
consequences of the method are taking the special significance, for which one 
does not need to know the exact basic functions $e_n(t)$, 
${\cal E}_n({\bf r})$.

\section*{3. EFFECTIVE NUMBER OF MODES}

So far as the mode number $n$ in general cannot be univalently associated with 
any other parameter of partial wave (except its energy), the only natural 
way to restore distribution of $u_n$, without solving (2), (7), is evaluation 
of nonlinear $k$-order moments of modal spectra 
$\sum_n u_n^k$. It can be done with use of iterated kernels theorem via 
sequential integration of functions $\Gamma_S({\bf r},{\bf r'})$ or 
$\Gamma_T(t,t')$. The moment of 1st order has a trivial meaning of total 
field energy
$$
U=\sum_nu_n=\int d^2r\,\Gamma_S ({\bf r},{\bf r})=\int dt\,\Gamma_T (t,t). 
$$

As it shown in Ref.~12, the higher moments determine a probability 
distribution of wave amplitude under conditions of wave mixing at a strong 
scatterer. The most important characteristic of field structure is an 
effective number of terms in decompositions (1), (5), 
(8)\cite{Gamo,Pasm,Starikov}, which is expressed through 2nd order moment 
$$
\begin{array}{c}
N_{\it eff}={\left( \sum_nu_n\right) ^2}
\bigg/\left( {\sum_nu_n^2}\right) \\
=U^2\,\bigg/\left( \int d^2r\int d^2r'
\,\left| \Gamma_S ({\bf r},{\bf r'})\right| ^2\right) 
=U^2\,\bigg/\left( \int dt\int dt'\,\left| \Gamma_T(t,t')\right| ^2\right).
\end{array}
\eqno(10)
$$
The value of $N_{\it eff}$ specifies the ability of the total 
field to produce interference effects between two arbitrary separate points 
of the beam cross-section\cite{Pasm,Leshchev} and changes from unity for 
spatial coherent one-mode wave to infinity for completely incoherent field.

Two equivalent forms of (10) reflect real interconnection 
between spatial and temporal parameters of a beam. If one determines effective
area of beam cross-section --- $S_{{\it eff}}$, area of spatial coherence
--- $\sigma _c$, pulse duration --- $T_{{\it eff}}$ and correlation time --- 
$\tau _c$ in the form 
$$
S_{{\it eff}}=\left( \int d^2r\,\Gamma_S ({\bf r},{\bf r})\right) ^2\,\bigg/
\left( \int d^2r\, \Gamma_S ^2({\bf r},{\bf r})\right),
\eqno(11.a) 
$$
$$
T_{{\it eff}}=\left( \int dt\,\Gamma_T (t,t)\right) ^2\,\bigg/\left( \int dt\,
\Gamma_T^2(t,t)\right),
\eqno(11.b) 
$$
$$
\sigma _c=\left( \int d^2r\int d^2\rho \,\left| 
\Gamma_S ({\bf r}+{\bf \rho}/2,{\bf r}-{\bf \rho}/2)\right| ^2\right) \, 
\bigg/\left( \int d^2r\,\Gamma_S^2({\bf r},{\bf r})\right),
\eqno(11.c) 
$$
$$
\tau _c=\left( \int dt\int d\tau \,\left| \Gamma_S (t+\tau /2,t-\tau/2)
\right|^2\right) \,\bigg/\left( \int dt\,\Gamma_S ^2(t,t)\right),
\eqno(11.d) 
$$
then relation (10) takes the form of equality for spatial and temporal
degrees of freedom of wavefield 
$$
\frac{S_{\it eff}}{\sigma _c}=\frac{T_{\it eff}}{\tau _c}. 
\eqno(12) 
$$
It means that {\sl number of coherence zones per beam cross-section 
is equal to number of different spatial patterns over the pulse duration}.

Three of introduced in (11) parameters --- $T_{\it eff}$, $S_{\it eff}$, 
$\sigma _c$ --- have quite traditional meaning\cite{Starikov} and 
need no special remarks. The averaged over beam cross-section coherence 
time $\tau _c$ describes time of global changing of field structure or, in 
other words, characteristic width by $t-t'$ of the degree of similarity 
$H(t,t')$ of spatial wave patterns. Definitions (10), (11) have no sensitivity 
to overall phase modulation of the field 
$$
E({\bf r},t)\iff E({\bf r},t)\exp \left( i\phi ({\bf r})+
i\psi (t)\right) 
$$
and, therefore, value of $\tau _c$ can rather significantly differ from local 
correlation time, which is defined in signal theory.

\section*{4. APPLICATION OF STATISTICAL AVERAGING}

Till this point the basic formalism has dealt with a wave-field 
$E({\bf r},t)$ as with the deterministic one. At the same time classical 
coherence theory usually operates with radiation 
characteristics, averaged over ensemble of similar fields, because in the 
majority of cases the individual pulse parameters are not of interest. 
Hence, the natural question arises --- how can such stochastic hypotheses 
influence on the results of previous analysis?

As a first step let us consider what statistical averaging gives at the 
stage of the basic integral equations (2), (7) formulation. Just as in the 
standard approach\cite{Gamo}, one can substitute the kernels 
$\Gamma_S ({\bf r},{\bf r'})$ and $\Gamma_T (t,t')$ with their 
averages $\left\langle \Gamma_S ({\bf r},{\bf r'})\right\rangle$ and 
$\left\langle\Gamma_T (t,t')\right\rangle$. However it is
evident, that transversal correlation function (1), averaged over time 
interval only, contains much more information about spatial structure of a 
beam, than similar value of 
$\left\langle \Gamma_S ({\bf r},{\bf r'})\right\rangle $ does. 
The approximate equality can take place only in the limit of infinite pulse 
duration and under quasiergodicity of the ensemble. Exactly the same with 
appropriate changing of words can be stated about temporal correlation function 
$\Gamma_T (t,t')$. It is easy to see that such a lack of information breaks
the main property of the present formalism --- its spatio-temporal duality.

As a result of this procedure the interpretation of the relations (1) -- (4) 
and (6) -- (9) must be changed. All other conclusions of the above 
sections (with exception of equality (12), which disappears) remain valid if 
taking into account that now we talk about two {\sl different and 
complementary} means for describing of coherence of random pulse ensemble 
(but not for a particular wave). For transversal coherence this will be 
nonstationary variant of Gamo's treatment\cite{Gamo} and for temporal 
correlation function it will have the form of modified Karhunen-Lo\'eve 
expansion\cite{Fu} with double averaging --- over ensemble and beam 
cross-section. In the subsequent discussion we shall utilise the fact that for 
each decomposition one can introduce its own number of degrees of freedom 
(10) --- spatial $N_S$ and temporal $N_T$ 
$$
N_S=\left\langle U^2\right\rangle \,\bigg/\left( \int d^2r\int d^2r' \,
\left| \left\langle \Gamma_S ({\bf r},{\bf r'})\right\rangle \right|^2\right)
=\lim _{N_T,T\rightarrow \infty }N_{\it eff},
\eqno(13.a) 
$$
$$
N_T=\left\langle U^2\right\rangle \,
\bigg/\left( \int dt\int dt'\,
\left| \left\langle \Gamma_T (t,t')\right\rangle \right| ^2\right)
=\lim _{N_S,S\rightarrow \infty }N_{\it eff}. 
\eqno(13.b) 
$$

Now it is clear that in order to preserve spatio-temporal duality of the 
formalism, the ensemble averaging should be applied at the later stages of 
consideration. As an example of such approach let us estimate the 
number of terms in the modal decomposition of a mean light pulse from the ensemble. The simplest and the most popular type of field statistics is 
Gaussian. In this case the coarse estimation can be done by averaging of 
integrals in definition of $N_{\it eff}$ (10), that, accounting the 
splitting of higher correlations and (13), gives a very simple formula 
$$
\overline{N_{\it eff}}=\frac{N_SN_T}{N_S+N_T}.
\eqno(14)
$$

On the basis of general reasons one can formulate some more requirements, 
which {\it a priori} should be satisfied by any admissible dependence 
$\overline{N_{\it eff}}(N_S, N_T)$. Thus, the function 
$\overline{N_{\it eff}}(N_S, N_T)$ must be symmetrical about permutation of 
its arguments because of the dual status of spatial and temporal degrees of 
freedom 
$$
\overline{N_{\it eff}}(N_S,N_T)=\overline{N_{\it eff}}(N_T,N_S). 
\eqno(15) 
$$
The value of $\overline{N_{\it eff}}$ must be a non-decreasing function of its 
arguments, that with (13) leads to conclusion
$$
\overline{N_{\it eff}} \le N_S, N_T
$$
and in asymptotics $N_S=const\cdot N_T\gg 1$ the average number of modes 
will be linear with respect to any of arguments, in particular,
$$
\overline{N_{\it eff}}(N_S=N_T=N\gg 1)\propto N. 
$$
Ensemble with only one degree of freedom in any of the basic subspaces 
corresponds to coherent (in terms of (5)) field 
$$
\overline{N_{\it eff}}(N_S=1,N_T)=\overline{N_{\it eff}}(N_S,N_T=1)=1. 
$$

It is easy to see that estimation (14) obeys all above requirements but the 
last one, i.~e. it poorly describes the region of small numbers of degrees of 
freedom (it is the consequence of approximate way of averaging 
$\langle N_{\it eff}\rangle$). The situation can be improved by taking into 
account the fluctuations not only of the denominator, but also of the numerator 
(energy of light pulses) of expression (10) 
$$
\left\langle U^2\right\rangle =\left\langle U\right\rangle ^2\left(1+
\varepsilon (N_S,N_T)\right),
$$
$$
\varepsilon =\frac 1{\left\langle U\right\rangle ^2}
\int d^2r\int d^2r'\int dt\int dt'
\left| \left\langle E({\bf r},t)E^{*}({\bf r'},t') \right\rangle \right| ^2.
$$
Correction $\varepsilon$ must satisfy the condition (15) and have the order
of magnitude $\varepsilon \propto 1/(N_S N_T)$, that can be confirmed by 
consideration cross-spectrally pure light\cite{Goodman}, when correlation 
function factorizes. Hence, the refined estimation of the number of modes 
in the mean pulse may be written as following
$$
\overline{N_{\it eff}}=\frac{N_SN_T+1}{N_S+N_T}.
\eqno(16) 
$$

For the first time estimation like (16) was given without a proof in 
paper\cite{Sidoro} for a system of several identical, statistically 
independent emitters with drifting phase. At limit $N_S,N_T\gg 1$ appropriate 
formula from Ref. 16 converts to (14). One can point out some more cases, 
which asymptotically lead to the same dependence. All this allows to say that 
area of applicability of relation (14) as estimation of 
$\overline N_{\it eff}$ is much wider than above assumptions.

In order to illustrate the consequences from relation (14) we can consider a 
very vivid example of the ensemble of Schell-model fields\cite{StWol,Desch} 
$$
\left\langle E({\bf r},t)E^{*}({\bf r'},t')\right\rangle =
\sqrt{I({\bf r},t) I({\bf r'},t')}\gamma ({\bf r}-{\bf r'},t-t'). 
\eqno(17) 
$$
One of the possible interpretations of model (17) corresponds to illustrative 
situation when a fast shutter cuts off a pulse of radiation from a primary 
steady-state uniform partially coherent source. Then just beyond the shutter 
the degree of coherence $\gamma ({\bf r} ,t)$ is specified by 
statistical parameters of the source only (say, with $\sigma$, $\tau$ being 
an area and a time of correlation, respectively), while $I({\bf r},t)$ is 
(within a factor) a deterministic function of the shutter transmittance 
($S$, $T$ --- the shutter aperture area and the time it is opened).
On substituting (17) into (13) and accounting (11), (14) one can assure that 
the effective area of coherence (in the mean pulse) depends not only on spatial 
parameters, but as well on ratio between temporal characteristics $T/\tau$ of 
the primary source and the shutter. And {\it vice versa}, the lifetime of a 
particular wavefront structure in the mean pulse is also determined by the 
ratio of $S/\sigma$. It explains the significance of the discussed modal 
formalism for the nonlinear optical and laser beam problems.

\section*{5. DISCUSSION}

In conclusion it is worth to point out the resemblance of 
the considered modal technique with bi-orthogonal decompositions used in other 
branches of physics --- e.~g. turbulence theory\cite{Lumley} and pattern 
recognition\cite{Fu}. Such tie is based on the common concept of complex 
process representation. By this analogy, the spatial partial coherence may be 
described as a sequence of more or less similar frames (instant wavefield 
structures) replacing each other. From this viewpoint coherent modes specify 
the feature basis of wavefronts evolution.

According to the general concept the application of global ensemble averaging 
procedure is efficient (it gives results with comparatively small relative 
variance) when the number of modes in the mean pulse is high. Nevertheless, 
there are situations where under small $\overline{N_{\it eff}}$ the number of 
the ensemble degrees of freedom in one of the subspaces is much more than it 
($N_T \gg \overline{N_{\it eff}}$ or $N_S \gg \overline{N_{\it eff}}$). In 
this case the statistical averaging over the corresponding complex 
substructure of the field may be useful.

How it is seen from (5), each single mode in deterministic decomposition 
produces factorized correlation function, i.~e. corresponds to cross-spectral 
pure light\cite{Goodman}. But if we go to the whole multimode field, the 
spectral purity vanishes. Moreover, for the statistical ensemble even 
one-mode field will, in general, not be cross-spectral pure.

Besides the discussed manifestations of spatio-temporal symmetry for modal 
decomposition, there exists a wide class of uncertainty 
relations\cite{Bastiaans}, where it must also appear. They should have the 
form of inequalities bounding modal characteristics with such parameters of 
wave as angular divergence and spectral bandwidth. This statement leans against 
the fact that proof of uncertainty relations for correlation functions does 
not depend on the type of averaging used.


\begin{thebibliography}{99}

\bibitem{Gamo}  H. Gamo, ``Matrix treatment of partial coherence'', 
{\it Progress in Optics}, Ed. E. Wolf, {\bf 3}, pp. 187-332, North-Holland,
Amsterdam, 1964

\bibitem{Wolf1}  E. Wolf, ``New theory of partial coherence in the 
space-frequency domain. Part I: Spectra and cross-spectra of steady-state 
sources'', {\it J. Opt. Soc. Am.}, {\bf 72}, pp.~343-351, 1982.

\bibitem{Wolf2}  E. Wolf, ``New theory of partial coherence in the 
space-frequency domain. Part II: Steady-state fields and higher-order 
correlations'', {\it J. Opt. Soc. Am.}, {\bf A3}, pp.~76-85, 1986.

\bibitem{DeSantis}  P. De~Santis, F. Gori, G. Guattary, C. Palma, 
``Synthesis of partially coherent fields'', {\it J. Opt. Soc. Am.}, 
{\bf A 3}, pp. 1258-1262, 1986.

\bibitem{Bastiaans}  M. J. Bastiaans, ``Uncertainty principle for partially 
coherent light'', {\it J. Opt. Soc. Am.}, {\bf 73}, pp. 251-255, 1983.

\bibitem{Pasm}  G. A. Pasmanik, V. G. Sidorovich, ``Interrelation between 
coherent properties of light beams and space-time structure'', {\it Radiophys. 
{\sl \&} Quantum Electron.}, {\bf 23}, pp. 809-814, 1980.

\bibitem{Lumley}  G. Berkooz, P. Holmes, J. L. Lumley, ``The proper
orthogonal decomposition in the analysis of turbulent flows'', {\it Ann. Rev.
Fluid Mech.}, {\bf 25}, pp.~539-575, 1993.

\bibitem{Lazaruk}  A. M. Lazaruk, ``Limits of applicability of the Raman-Nath 
approximation in problems of radiation self-diffraction'', 
{\it Optics {\sl \&} Spectroscopy (USA)}, {\bf 53}, pp.~633-636, 1982.

\bibitem{ZeldShkun}  B. Ya. Zel'dovich, N. F. Pilipetsky, V. G. Shkunov, 
{\it Principles of phase conjugation}, Springer-Verlag, Berlin, 1985.

\bibitem{StWol}  A. Starikov, E. Wolf, ``Coherent-mode representation of 
Gaussian Schell-model sources and of their radiation fields'', 
{\it J. Opt. Soc. Am.}, {\bf 72}, pp. 923-928, 1982.

\bibitem{Gori}  F. Gori, G. Guattary, C. Padovani, ``Modal expansion for 
$J_0$-correlated Schell-model sources'', 
{\it Opt. Comm.}, {\bf 64}, pp.~311-316, 1987.

\bibitem{Leshchev}  A. A. Leshchev, ``Method for measuring the number of 
spatially coherent modes in optical radiation'', 
{\it Optics {\sl \&} Spectroscopy (USA)}, {\bf 55}, pp.~599-600, 1983

\bibitem{Starikov}  A. Starikov, ``Effective number of degrees of freedom 
of partially coherent sources'', 
{\it J. Opt. Soc. Am.}, {\bf 72}, pp.~1538-1544, 1982.

\bibitem{Fu}  Fu K. S. {\it Sequential methods in pattern recognition and
machine learning}, Academic Press, New York -- London, 1968.

\bibitem{Goodman}  J. W. Goodman, {\it Statistical Optics}, A Wiley-%
Interscience Publication, New York, 1985.

\bibitem{Sidoro}  M. V. Vasil'ev, V. Yu. Venedictov, A.~A.~Leshchev, 
P.~M.~Semenov, V.~G.~Sidorovich, O.~V.~Solodyankin, ``Reduction of 
speckle-structure contrast in image under laser illumination'', 
{\it Optics {\sl \&} Spectroscopy (USA)}, {\bf 70}, pp.~6-7, 1991

\bibitem{Desch}  J. Deschampus, D. Courjon, J. Bulabois, ``Gaussian 
Schell-model sources: an example and some perspectives'', 
{\it J. Opt. Soc. Am.}, {\bf 73}, pp.~256-261, 1983

\end{thebibliography}
\end{document}